\documentstyle[12pt]{article}

\begin{document}
\baselineskip = 24pt

\begin{titlepage}

\hoffset = .5truecm
\voffset = -2truecm

\centering

\null
\vskip -1truecm
\rightline{IC/95/}
\vskip 1truecm

{\normalsize \sf \bf International Atomic Energy Agency\\
and\\
United Nations Educational, Scientific and Cultural Organization\\}
\vskip 1truecm
{\huge \bf
INTERNATIONAL CENTRE\\
FOR\\
THEORETICAL PHYSICS\\}
\vskip 3truecm

{\LARGE \bf
Determinant of twisted \\chiral Dirac operator on the lattice.
\\}
\vskip 1truecm

{\large \bf
C. D. Fosco
\\}
\medskip
{\large      and\\}
\medskip
{\large \bf
S. Randjbar-Daemi
\\}

\vskip 6truecm

{\bf International Centre for Theoretical Physics \\}
April 1995
\end{titlepage}

\hoffset = -1truecm
\voffset = -2truecm

\title{\bf
Determinant of twisted \\chiral Dirac operator on the lattice.}
\vspace{2cm}

\author{
{\bf
C. D. Fosco}\\ \\
{\normalsize and}\\
\normalsize
{\bf}\\
{\bf S. Randjbar-Daemi}\\ \\
\normalsize International Centre for Theoretical Physics, Trieste 34100,
{\bf Italy} }

\date{May 1995}
\newpage

\maketitle

\begin{abstract}
Using the overlap formulation,
we calculate the fermionic determinant on the lattice for chiral fermions
with twisted boundary conditions in two dimensions.
When the lattice spacing tends to zero we recover the results of the usual
string-theory continuum calculations.
\end{abstract}

\newpage


The construction of a viable lattice regularization for chiral gauge
theories has
been the subject of renewed interest and research, since the appearance of
the overlap formalism~\cite{nar}. This is a new
proposal which is based on an earlier idea of Kaplan~\cite{kap} and seems
to bypass the
kinematical constraint stated by the Nielsen-Ninomiya
theorem~\cite{kn:Karsten.}. This method defines the determinant of
a chiral Dirac operator in $2 d$ dimensions as an overlap between the
Dirac vacuum states of two auxiliary Hamiltonians  acting on
Dirac fermions in $2 d + 1$ dimensions.

There are two issues that require verification if the overlap
formulation is to be accepted as a procedure to regulate chiral gauge
theories. Firstly, one must check in the continuum whether the overlap
yields the right determinant for any chiral Dirac operator, reproducing
their well-known distinctive properties: anomalies, zero-modes in non-trivial
backgrounds, etc. In a Hamiltonian approach these continuum tests
have been performed in $2$ and $4$ dimensions in references ~\cite{nn1}
and ~\cite{ran1}, respectively. The $4$-dimensional continuum results have
also been confirmed  in a $5$-dimensional approach in ~\cite{chin}.
Secondly, one has to show that the lattice version of the overlap does not
suffer from any of the drawbacks  that afflicted previous attempts to
regulate fermions on the lattice. Analytic lattice calculations have been
carried out only for slowly varying background gauge fields in the
Schwinger Model in ~\cite{aoki}
and in non-Abelian gauge theories in $4$ dimensions in [10]. There
are also numerical calculations which confirm the overlap picture [1, 11].
In this letter we check the overlap for
the  particular situation of fermions with twisted boundary conditions
on a discretized torus. This is
a case for which the overlap can be  analytically calculated, even on the
lattice.
Our results reinforce the conclusions of \cite{twi} where this problem
was studied numerically.

Recently, some other approaches to the problem of chiral fermions on
the lattice have been proposed~\cite{sla,tho}, which will not be
studied here.

We review first the case of chiral fermions on a two-dimensional
continuum torus $T^2$, which is
described in terms of two real coordinates $\sigma^1$, $\sigma^2$ lying
in the range $0 \leq \sigma^{\mu} < 1$, and equiped with the constant
Euclidean metric:
\begin{equation}
ds^2 = \mid d \sigma^1 \, + \, \tau d \sigma^2 \mid^2 \,=\,
g_{\mu \nu} \, d \sigma^{\mu} \, d \sigma^{\nu} \;\;, \;\; {\rm Im}\,\,\tau
> 0 \;\;.
\label{2.1}
\end{equation}
The free chiral Dirac operator $D$ on $T^2$ is then given by
$D \;=\; \gamma^a e^{\mu}_a \partial_{\mu} \, \displaystyle{{1+
\gamma_5}\over{2}}$,
where the zweibeins $e^{\mu}_a$ are: $e^{\mu}_1 = (1,0)$,
$e^{\mu}_2 = (-\frac{\tau_1}{\tau_2} , \frac{1}{\tau_2})$,
and the Dirac matrices are chosen to be: $\gamma^1 = \sigma_1$,
$\gamma^2 = \sigma_2$, $\gamma^5 = \sigma_3$ with $\sigma_a$ the
usual Pauli matrices.

The chiral fermionic field $\psi (\sigma)$ is subject to the
twisted boundary conditions
\begin{equation}
\psi (\sigma^1 + 1, \sigma^2) \;=\, - e^{i \phi_1} \;\psi (\sigma^1,\sigma^2)
\;\;,\;\;
\psi (\sigma^1, \sigma^2 + 1) \;=\, - e^{i \phi_2} \; \psi (\sigma^1,\sigma^2)
\label{2.3}
\end{equation}
where $\phi_1$, $\phi_2$ are real numbers, with values between $0$ and
$2 \pi$.
The determinant of $D$, evaluated for fields satisfying
(\ref{2.3}) is equivalent to the one of fermions with
antiperiodic boundary conditions in both directions, on the background of
a constant-field $A_{\mu} \,=\, \phi_{\mu}$, i.e., the determinant of
$D(A) \;=\; \gamma^a e^{\mu}_a (\partial_{\mu} + i A_{\mu})\,
\displaystyle{{1+\gamma_5}\over{2}}$.
The result for the
normalized determinant, taken from ~\cite{alv}, but adapted to our
conventions, is:
\begin{equation}
\frac{\det D(A)}{\det D(0)} \;=\; e^{-\frac{i}{2 \pi} A_1 A_2}\;
q^{\frac{1}{8 \pi^2}A_1^2} \; \frac{\vartheta(\alpha,\tau)}{
\vartheta( 0 ,\tau)}\;,
\label{2.4}
\end{equation}
where $\vartheta (\alpha, \tau) \;=\; \sum_{n = -\infty}^{n=+\infty}
e^{i \pi \tau n^2 + i 2 \pi n \alpha}$, $\alpha =
\frac{1}{2\pi} (\tau A_1 - A_2)$ and $q = e^{2 \pi i \tau}$.
To calculate the determinant of $D(A)$ on the lattice, we follow
the general procedure outlined in \cite{ran}. We define a two-component
fermionic field $\Psi (n)$ with $n \,=\,(n^1,n^2)$ on an $N \times N$
square lattice:
$1\leq n^1 \leq N$, $1 \leq n^2 \leq N$.
With the restriction $0 \leq \sigma_{\mu} \leq 1$, the lattice spacing is
$N^{-1}$. In this way we recover a torus of the proper size
in the continuum limit ($N \to \infty$).
To define the overlap, two second-quantized Dirac Hamiltonians
$H_{\pm}(A)$ are introduced:
\begin{equation}
H_{\pm}(A) \;=\; \sum_{m,n} \frac{1}{\tau_2} \;\Psi^{\dagger} (m) \,
{\cal H}_{\pm}^{(0)} (m-n) \, U(m,n) \, \Psi (n) \;\;,
\label{2.10}
\end{equation}
where the link variables $U(m,n) \,=\, \exp [ \frac{i}{N} (m-n)
\cdot A ]$,
and ${\cal H}_{\pm}^{(0)}$ are the respective, free, one-body Hamiltonians
on the lattice. Their Fourier-space representations are:
\begin{equation}
\tilde{\cal H}_{\pm}^{(0)} (k) \;=\; \gamma_5 \left[ \;i \gamma^a
e^{\mu}_a C_{\mu} (a k)\,+\, B(ak) \,
\pm \, \mid \Lambda \mid \;\right] \;,
\label{2.11}
\end{equation}
where $C_{\mu} (p) = \sin (p_{\mu})$, $a \,=\,\frac{2 \pi}{N}$, $B$ is
a function  which satisfies the conditions:
\begin{eqnarray}
\mid p \mid << 1 & &\Rightarrow B(p) \sim r p^2 \;\;, \nonumber\\
p \neq 0 \,,\, C_{\mu}(p) = 0 & & \Rightarrow
B(p)^2 > \Lambda^2 \;\;,
\label{2.12}
\end{eqnarray}
in order to eliminate the doublers~\cite{ran}, and $\Lambda$ is
a constant with the dimension of a mass.
$H_{\pm}$ are diagonal in momentum space
\begin{equation}
H_{\pm}(A) \;=\; \sum_{k} {\tilde{\Psi}}^{\dagger} (k)\,
{\tilde{\cal H}}_{\pm} (k\mid A) \, {\tilde{\Psi}}(k)\;,
\end{equation}
where:
\begin{equation}
{\tilde{\cal H}}_{\pm} (k \mid A) \;=\;
{\tilde{\cal H}}^{(0)} (k + \frac{A}{2 \pi})\;.
\label{2.13}
\end{equation}
and $k_{\mu}$ takes half-integer values, which are summed in (7) over
the range $[- \frac{N}{2} , \frac{N}{2} ]$.

The normalized chiral determinant in the presence of the external field,
is represented in the overlap formalism by:
\begin{equation}
\frac{\det D(A)}{\det D(0)}\,=\,
\frac{\langle + \mid A+ \rangle}{\mid \langle + \mid A+ \rangle \mid}
\frac{\langle A+ \mid A- \rangle}{\langle + \mid - \rangle}
\frac{\langle A- \mid - \rangle}{\mid \langle A- \mid - \rangle \mid}
\label{2.135}
\end{equation}
where $\mid A\pm \rangle$ denote the Dirac vacua for the
Hamiltonians $H_{\pm}(A)$, respectively; and $\mid \pm \rangle
\,=\,\mid 0 \pm \rangle$.
In terms of the expansion of $\Psi$ in eigenspinors of the
one-body Dirac Hamiltonians
\begin{equation}
\Psi(n) \;=\; \frac{1}{\tau_2} \sum_k \, \left \{  b_{\pm} (k\mid A)
u_{\pm} (k \mid A) \;+\; d^{\dagger}_{\pm} (k \mid A)
v_{\pm} (k \mid A)\right \}  \, e^{ i 2 \pi k \cdot n} \;\;,
\label{2.14}
\end{equation}
(\ref{2.135}) is given by
\begin{equation}
\frac{\det D(A)}{\det D (0)} \,=\, \prod_k
\left \{ \,\frac{ v^{\dagger}_+(k\mid 0) v_+(k \mid A)}
{ \mid v^{\dagger}_+ (k\mid 0) v_+(k \mid A) \mid }
\frac{ v^{\dagger}_+(k\mid A) v_-(k \mid A)}
{v^{\dagger}_+(k\mid 0) v_-(k \mid 0)}
\frac{ v^{\dagger}_-(k\mid A) v_-(k \mid 0)}
{\mid v^{\dagger}_-(k\mid A) v_-(k \mid 0)\mid} \,\right \} \;,
\label{2.15}
\end{equation}
where the dependence on the momentum and the value of the constant
field has been explicitly indicated.
By virtue of (\ref{2.13}), all the eigenspinors $v_{\pm}(k\mid A)$,
which are the negative-energy eigenvectors of
${\cal H}_{\pm}(k\mid A)$ in (\ref{2.15}),
can be  obtained from the free ones:
\begin{eqnarray}
v_+ (k) &=&
[2 \gamma_+ (k)]^{-\frac{1}{2}}
\left( \begin{array}{c}
\displaystyle{\frac{{\bar \tau} C_1 (a k) - C_2 (a k)}{\sqrt{\gamma_+ (k)
+ b(k) +\mid \lambda \mid}}} \\ \\
\sqrt{\gamma_+ (k) + b (k) + \mid \lambda \mid}
\end{array} \right) \;, \nonumber\\ & &\nonumber\\
v_- (k) &=&
[2 \gamma_-(k)]^{-\frac{1}{2}}
\left( \begin{array}{c}
\sqrt{\gamma_- (k) - b (k) + \mid \lambda \mid} \\ \\
\displaystyle{\frac{ \tau C_1 (a k) - C_2 (a k)}{\sqrt{\gamma_-
(k) - b(k) +\mid \lambda \mid}}}
\end{array} \right)
\label{sp1}
\end{eqnarray}
where
\begin{equation}
\gamma_{\pm} (k) \,=\, \sqrt{ (b(k) \pm \mid \lambda \mid)^2 \,+\,
\mid   C_2 (a k) \,- \, \tau C_1
(a k) \mid^2 } \;,\;
b(k) \,=\, \tau_2 \,\, B ( a k) \; , \;
\lambda \,=\, \tau_2 \Lambda \;\;,
\label{2.17}
\end{equation}

The result is:
\begin{equation}
\frac{ \det D (A) }{\det D (0)} \;=\; \prod_k
\,f(k,A)\,e^{i g(k,A)}\, \left[ \, \frac{\tau C_1 (\, a\, (k+\frac{A}{2\pi}
\, ) \,)\, - \,
 C_2 (\, a \, (k+\frac{A}{2\pi}\, )\, )}{ \tau C_1 (ak) - C_2 (a k)} \,
\right]
\label{2.16}
\end{equation}
where
\begin{eqnarray}
f(k,A) &=&
\left[ \frac{\gamma_+ (k + \frac{A}{2 \pi} ) +
\gamma_- (k + \frac{A}{2 \pi} ) + 2 \mid \lambda \mid}{ \gamma_+ (k) +
\gamma_- (k) + 2 \mid \lambda \mid} \right]
\, \times \, \left[ \frac{\gamma_+ (k) \gamma_- (k)}{\gamma_+ (k+
\frac{A}{2\pi})
\gamma_- (k + \frac{A}{2 \pi} )} \right. \nonumber\\
& & \left. \frac{\gamma_+(k) + b(k) + \mid \lambda \mid}{\gamma_+
(k+\frac{A}{2\pi}) + b (k+\frac{A}{2\pi}) +
\mid \lambda \mid}
\frac{\gamma_-(k) - b(k) + \mid \lambda \mid }{\gamma_-
(k+\frac{A}{2\pi}) - b (k+ \frac{A}{2\pi})
+ \mid \lambda \mid} \right]^{\frac{1}{2}}
\label{2.18}
\end{eqnarray}
and the phase $g(k,A)$ is equal to the argument of the complex number $z$:
\begin{eqnarray}
z &=& z_+ \;\; z_- \nonumber\\
z_+ &=& [\, \gamma_+ (k) + b(k) + \mid \lambda \mid \,]\;\;
[\, \gamma_+ (k+\frac{A}{2\pi}) + b(k+\frac{A}{2\pi}) + \mid \lambda \mid \,]
\nonumber\\
& & +\; [\, \tau C_1 (a k) - C_2(a k) \,] \;\;
 [\, {\bar \tau} C_1 (\,a (k+\frac{A}{2 \pi})\,) - C_2(\,a( k+
\frac{A}{2 \pi})\, )\, ]
\nonumber\\
z_- &=& [\, \gamma_- (k) - b(k) + \mid \lambda \mid \,] \;\;
[\, \gamma_- (k+\frac{A}{2\pi}) - b(k+\frac{A}{2\pi}) + \mid \lambda \mid \,]
\nonumber\\ & & + \; [ \, \tau C_1 (a k) - C_2(a k) \, ]
 \;\;[ \,{\bar \tau} C_1 (\,a (k+\frac{A}{2 \pi})\,) -
C_2(\, a (k+\frac{A}{2 \pi})\, )\, ]  \;.
\label{ph}
\end{eqnarray}
We shall consider the effective action $\Gamma$
\begin{equation}
\Gamma (A) \;=\; - \, \log \left[ \frac{ \det D (A) }{ \det D (0) } \right]
\;,
\label{rb1}
\end{equation}
in the continuum limit, which in our conventions (i.e., $a =
\frac{2 \pi}{N}$) is tantamount to taking $N \to \infty$.
In taking this limit, an expansion will
be used to get the analytic result for $\Gamma$, depending on
the values of the momenta $k_{\mu}$ involved in the infinite product
(\ref{2.16}). The contribution coming from the low-momenta, denoted
$\Gamma_{hol}$ for reasons that will become clear below, is obtained
by assuming  $a k_{\mu} << 1$
(note that the  maximum possible value for this number is $\pi$).
As $N$ is large, and $\frac{A}{2 \pi} \leq 1$,  $a \,\frac{A}{2 \pi}$
is also very small. However, we do not consider $k >> \frac{A}{2 \pi}$
in the present expansion (that case will be discussed later on).
Under the above-mentioned  conditions, the functions $C_{\mu}$ can be
expanded near zero:
\begin{equation}
C_{\mu} (a \, (k + \frac{A}{2 \pi}) \,) \, \sim \, a \,(k +
\frac{A}{2 \pi})_{\mu}
\;\; , \;\; C_{\mu} (a k ) \, \sim \, a k_{\mu}  \;,
\label{rb2}
\end{equation}
but the periodicity of $C_{\mu}$ on the momentum torus implies that
there are other points in momentum space (the doublers) giving contributions
of the same order. This is the manifestation of the doubling problem in
this simple context. The unwanted contributions are eliminated by a
careful choice of $B$ and $\Lambda$. We first note that for the
normal zero of $C_{\mu}$, the functions $f$ and $g$ play no role
(i.e., $f \to 1$ and $g \to 0$), if we require~\cite{ran}:
\begin{equation}
\Lambda^2 >> a^2 k^2 \;\;, \;\;  \Lambda^2 >> B^2 \;,
\label{rb3}
\end{equation}
for momenta near this zero, since (\ref{rb3}) implies that $\gamma_{\pm}
\, \sim \, \mid \lambda \mid$. Thus for the normal zero the only contribution
to $\Gamma_{hol}$ will come from (the expansion near zero of) the
term in square brackets in (\ref{2.16}).

When the momenta approach  one of the doublers, we
get instead:
\begin{equation}
\gamma_{\pm} \,\sim \, \mid b \, \pm \, \mid \lambda \mid \,\mid \; +\;
\frac{1}{2} \frac{\mid \tau C_1 - C_2 \mid^2}{\mid \, b \pm \mid \lambda
\mid \,\mid} \;.
\label{rb4}
\end{equation}
and then assuming $B^2 >> \Lambda^2$ for these momenta, one
sees that all but one factor in $f(k,A)$ tend to one, the relevant
contribution being given by:
\begin{equation}
f \, \sim \, \sqrt{ \frac{\gamma_-(k) - b(k) + \mid \lambda \mid }{\gamma_-
(k+\frac{A}{2\pi}) - b (k+ \frac{A}{2\pi}) + \mid \lambda \mid} } \, \sim \,
\frac{\mid \tau \,C_1 (ak) - C_2 (a k)\mid}
{\mid \tau C_1 (a(k+\frac{A}{2\pi})) - C_2 (a (k+
\frac{A}{2\pi}))\mid} \; ,
\label{rb5}
\end{equation}
whereas regarding the phase $g(k, A)$, $z_+$ becomes real in the limit,
and $z_-$ yields a non-zero contribution
\begin{eqnarray}
g (k,A) & \sim & {\rm arg} z_- \;\sim \; {\rm arg} \left( [ \, \tau C_1 (a k)
- C_2(a k) \, ]
 \;[ \,{\bar \tau} C_1 (\,a (k+\frac{A}{2 \pi})\,) -
C_2(\, a (k+\frac{A}{2 \pi})\, )\, ] \right) \nonumber\\
 & = & {\rm arg} [ \tau C_1 (a k) - C_2 (a k) ] \,- \,
{\rm arg} [ \tau C_1 (a \, (k + \frac{A}{2 \pi}) \,) - C_2 (a \,(k +
\frac{A}{2 \pi})\, )] \;.
\label{rb6}
\end{eqnarray}
Putting together (\ref{rb5}) and (\ref{rb6}), we realize that they
exactly cancel the corresponding contribution of the doubler coming
from the factor in square brackets
in (\ref{2.16}). So we have seen that it is  only necessary to expand near the
trivial zero on $C_{\mu}$, and the corresponding expression
for $\Gamma_{hol}$ becomes:
\begin{equation}
\Gamma_{hol} (A) \;=\; - \, \lim_{N\to \infty} \sum_k
\; \log \frac{\tau k_1 - k_2 + \alpha}{\tau k_1 - k_2} \;,
\label{a1}
\end{equation}
where the summation over $k$ now ranges over half-integers from
$- \infty$ to $+ \infty$ for both components.
This expression (\ref{a1}) is already explicitly holomorphic
in the variable  $\alpha$, since it only depends on the gauge fields
in that particular combination.
We calculate it from (\ref{a1}) by rewriting the logarithm as
\begin{equation}
\log \frac{ \tau k_1 - k_2 + \alpha}{\tau k_1 - k_2}
\;=\; \int_0^{\tau k_1 + \alpha}
\frac{dt}{t - k_2} \,-\, \int_0^{\tau k_1} \frac{d t}{t - k_2} \;,
\label{2.20}
\end{equation}
then the summation over $k_2$
can be performed exactly, yielding
\begin{eqnarray}
\Gamma_{hol} (A) &=& + \sum_{k_1} \, \int_{\pi \tau k_1}^{\pi (\tau k_1
+ \alpha) } \, dt \, \tan t \nonumber\\
&=& - \, \sum_{k_1} \, \log \left[ \, \frac{ \cos ( \pi ( \tau k_1 +
\alpha ) )}{ \cos ( \pi \tau k_1 )} \right] \;.
\label{ex}
\end{eqnarray}
After writing the cosines in terms of exponentials, one can cancel factors
in numerator and denominator
and then use the infinite-product representations of $\vartheta$-functions,
to obtain:
\begin{equation}
\Gamma_{hol} (A) \;=\; - \, \log \frac{ \vartheta (\alpha, \tau)}{
\vartheta (0,\tau)} \; .
\label{2.21}
\end{equation}
It is worth mentioning the role played by the lattice in the
definition on a precise procedure to handle the series over $k_1$ and $k_2$,
since the fact that one takes the {\em symmetric} limit for the sum over $k_2$
allows one to pair together terms with positive and negative values of
$k_2$, which renders the series convergent. Note also that, had the
doublers not been suppressed, we would have had to expand also near
the extra zeroes introducing the unwanted chiralities.

The behaviour of $\Gamma_{hol}$ under the discrete group of large
gauge transformations: $A_{\mu} \to A_{\mu} + 2 \pi n_{\mu}$ where
$n_{\mu}$ are integers, or equivalently: $\alpha \to \alpha + \tau
n_1 - n_2$, is simply obtained from the well-known properties of
$\vartheta$-functions, yielding:
\begin{equation}
\Gamma_{hol} \; \to \; \Gamma_{hol} \,+\, i \,
( \pi \tau n_1^2 \, + \, 2 \pi n_1 \alpha ) \;.
\end{equation}
This transformation law for $\Gamma_{hol}$ spoils the gauge
invariance of $\Gamma$, not only in its imaginary part (which
must change because of the anomaly), but also
of the real part. However, in Equation (14) one can check that
the real part of the regularized (finite $N$) effective action is
indeed gauge invariant. The breaking of gauge invariance of the
real part is a consequence of the approximation used
in deriving (\ref{2.21}) which fails to take into
account the high-momentum modes, for which the expansion
used for the functions $C_{\mu}$ breaks down. The effect of high momentum
modes is the introduction of a counterterm quadratic in $A_{\mu}$,
as power-counting shows.  This  holomorphic anomaly
exists also in the continuum calculations where the formal chiral
determinant is explicitly holomorphic, but a non-holomorphic piece
is introduced through the regularization. This contribution is
regularization-dependent, so its actual value can be modified
by the addition of (finite) quadratic counterterms. The
arbitrariness involved in deriving the non-holomorphic piece, can
for example be seen in that the separation between the low and large
momentum regimes is not a clear-cut one.

The real part of this quadratic counterterm is however  completely fixed
in our case by requiring its variation to cancel the real part of
(27), since the regularization (14) preserves gauge-invariance of
the modulus of the chiral determinant.
The imaginary part of this quadratic counterterm is fixed, to compare
with \cite{alv}, by  adding to it the appropriate global phase factor,
obtaining
\begin{equation}
\Gamma_{anom} (A) \;=\; + \frac{i}{2 \pi} A_1 A_2 \,-\,
\frac{i \tau}{4 \pi} A_1^2 .
\label{2.23}
\end{equation}
We have verified that the continuum overlap reproduces the real
part of (\ref{2.23}) exactly.

The full effective action $\Gamma$, given by the sum of the
holomorphic and anomalous parts should then read as
\begin{equation}
\Gamma (A) \;=\; \Gamma_{hol} (A) \,+\, \Gamma_{anom} (A) \;=\; - \,
\log \frac{ \vartheta (\alpha, \tau)}{ \vartheta (0,\tau)} \;
+ \;\frac{i}{2 \pi} A_1 A_2 \,-\,
\frac{i \tau}{4 \pi} A_1^2 \;,
\label{result}
\end{equation}
which agrees with the result of ~\cite{alv} exactly.
The numerical
results of  ~\cite{twi} are compatible with the exact result
of ~\cite{alv}, and hence with ours.

The variation of $\Gamma_{anom}$ under the large gauge transformations
is
\begin{equation}
\Gamma_{anom} \; \to \; \Gamma_{anom} \,-\, i ( \pi \tau n_1^2 \,+\,
2 \pi n_1 \alpha \,-\, A_1 n_2 ) \;,
\end{equation}
hence the variation of $\Gamma$ is simply:
\begin{equation}
\Gamma \;\to \; \Gamma \,+\, i A_1 n_2 \; ,
\end{equation}
i.e., linear in the fields and the group parameters.

\section*{Acknowledgements.}
We are appreciative of useful discussions with  Prof. R. Iengo
and Prof. J. Strathdee. We also acknowledge Prof. H. Neuberger
for pointing out an error in an earlier version of this letter.

\end{document}